\documentclass[nofootinbib,onecolumn]{revtex4}
\usepackage{graphicx}
\usepackage{subfig}
\usepackage{amssymb}
\usepackage{amsmath}
\usepackage{bm}
\usepackage{slashbox}
\usepackage{mathtools}
\usepackage{xcolor}

\begin{document}
\title{Global monopole as a generator of a bulk-brane structure in Brans-Dicke bulk gravity}
\author{Thiago R. P. Caram\^es}
\email{trpcarames@gmail.com}
\affiliation{
Departamento de F\'isica,
Universidade Federal de Lavras (UFLA)\\
Lavras-MG, Brazil}
\author{J. M. Hoff da Silva}
\email{julio.hoff@unesp.br}
\affiliation{Departamento de F\'isica e Qu\'imica, Universidade Estadual Paulista (UNESP),\\
 Guaratinguet\'a-SP, Brazil}

\pacs{}

\begin{abstract} 
We investigate a braneworld model generated by a global monopole in the context of Brans-Dicke gravity. After solving the dynamical equations we found a model capable to alleviate the so-called hierarchy problem. The obtained framework is described by a hybrid compactification scheme endowed with a seven-dimensional spacetime, in which the brane has four non-compact dimensions and two curled extra dimensions. The relevant aspects of the resulting model are studied and the requirements to avoid the well known seesaw-like behavior are discussed. We show that under certain conditions it is possible to circumvent such a pathological behavior that characterizes most of the models that exhibit hybrid compactification. Lastly, we deepen our analysis by considering possible extensions of this model to a setup with multiple branes and orbifold-like extra dimension. For this, we compute the consistency conditions to be obeyed by this more general configuration as predicted by the braneworld sum rules formalism. This study indicates the possibility of exclusively positive brane tensions in the model.       
\end{abstract}

\maketitle
\noindent

\section{Introduction}

In 1999 it was introduced a paradigmatic extra dimensional model of universe \cite{RSI} whose geometrical/topological setup makes possible to approach the hierarchy problem. Since then a great amount of work was done in many contexts within the scope of extra dimensions. The model presented in Ref. \cite{RSI} is built upon the $\mathbb{R}^4\times S^1/\mathbb{Z}_2$ configuration and, as such, may be faced as a type of effective braneworld model of the Horava-Witten \cite{HW} investigation relating the eleven-dimensional orbifold $\mathbb{R}^{10} \times S^1/\mathbb{Z}_2$ with ten-dimensional heterotic strings. 

From the gravitational point of view, the branes presented in Ref. \cite{RSI} are nothing but domain walls, one of them performing the large scale experienced Universe, with four non-compact dimensions. It became natural to wonder if other topological defects could be used to envisage a braneworld model with physical appealing characteristics. In this regard a branch of research was pursued by means of using different versions of cosmic strings to generate the bulk/brane structure. Global topological defects were used, leading to interesting braneworld models \cite{global1,global2,glo}. On the other hand, it is a well known fact that string theory encompasses a dilatonic scalar field in the low energy limit of the gravity \cite{ST}. The Brans-Dicke theory \cite{BD} appears in this context as the simplest theory in which the gravitational phenomena may be described by both a dilatonic field and a spin two particle. Therefore it may be faced as a good laboratory to explore extra dimensional more involved scenarios. This program was partially covered in the past. In Ref. \cite{local} a local cosmic string generating the braneworld model was studied, while in Ref. \cite{global} a braneworld was investigated with the aid of a global cosmic string. In both cases a six-dimensional bulk endowed with a five-dimensional brane arose and, therefore, the resulting models falls within the hybrid compactification scheme. In the present work we pursue the idea of using another topological defect, the global monopole, engendering the bulk/brane structure. After solving analytically the Brans-Dicke dynamical equations in the Einstein’s frame, we arrive at a hybrid compactification model with a seven-dimensional spacetime whose brane has four non-compact dimensions and two curled extra dimensions.

A global monopole is a topological defect resulting from the breaking of a global $SO(3)$ symmetry, whose gravitational effects have been first investigated by Barriola and Vilenkin within the context of General Relativity (GR) in \cite{barriola}. In \cite{lousto} Harari and Loust\'o went beyond and studied the consequences arising when the mass of the global monopole comes into play. They obtained a negative sign for this parameter and interpreted this result as an emergence of a repulsive gravitational potential around the defect.

In the present work we pursue the idea of using a global monopole to engender the bulk/brane structure. After solving analytically the Brans-Dicke dynamical equations in the Einstein's frame, we arrive at a hybrid compactification model with a seven-dimensional spacetime whose brane has four non-compact dimensions and two curled extra dimensions. We present a qualitative analysis contrasting the result obtained with its usefulness in serving as a braneworld model. The net result is that the model has an interesting setting, allowing an approach to the hierarchy problem while the size of the on-brane extra dimensions are in agreement with current experiments, thus avoiding a {\it seesaw-like} behavior \cite{global,el} \footnote{In some models   these two results are not achieved at same time: when the hierarcy problem is alleviated, the size of the extra dimensions evades the scale allowed by the observations and vice-versa. When this occurs one says that the model suffers with a seesaw-like behavior.}. The scalar field shape, however, is not free of ill defined behavior along the transverse extra dimension and this fact motivates our further study of braneworld models based on global monopole by means of the so-called braneworld sum rules. Applied to the case at hands we see that the sum rules point to the possibility of exclusively positive brane tensions constituting the model. While this last fact is shared by other models, and in this sense is not completely new, this is certainly a piece of good news, since a negative brane tension is a necessity in the paradigmatic model \cite{RSI}. 

This paper is organized as follows. In the Section II we introduce the basic ingredients used to construct the desired framework, namely, a braneworld model based on a global monopole. This section of composed of two subsections: IIA, where we explicitly construct an analytic solution and IIB, devote to a discussion of the qualitative aspects of the obtained solution showing both the virtues and setbacks in engendering an alternative braneworld model. In the Section III we take one step further in our analysis by examining possible extensions via braneworld sum rules. Finally, in the Section IV we summarize the main results of this work and leave our concluding remarks. 

\section{Towards The model}

Let us assume a brane-bulk structure generated by a $(p+4)$-dimensional global monopole within the Brans-Dicke gravity. The symmetry obeyed by such a configuration leads us to start from the spherically symmetric line of element below   
\begin{equation}
ds^2=e^{2\gamma(r)}dt^2-e^{2\alpha(r)}dr^2-e^{2\beta(r)}(d\theta^2+\sin^{2}\theta d\phi^2)-e^{2\sigma(r)}\sum^{p}_{i=1}dz_{i}^{2}.
\end{equation}
The corresponding field equations in the Einstein's frame (where the metric and the dilaton are decoupled) are
\begin{equation}
\label{feq1}
R_{\nu}^{\mu}=2 \partial^{\mu}\varphi \partial_{\nu}\varphi + 8\pi {\cal G}\left(T_{\nu}^{\mu}-\frac{\delta_{\nu}^{\mu}T}{p+2}\right)
\end{equation}
and
\begin{equation}
\label{feq2}
\Box \varphi = -4 \pi {\cal G} \tilde{\alpha} T,  
\end{equation}
where ${\cal G}$ is the gravitational coupling constant and the parameter $\tilde{\alpha}$ is combination involving the Brans-Dicke one given by $\tilde{\alpha}^2=1/(2\omega+3)$. The conservation law of the energy-momentum tensor reads
\begin{equation}
\label{consL}
\nabla_{\mu}T_{\nu}^{\mu}=\tilde{\alpha}T\partial_{\nu}\varphi. 
\end{equation}
On the assumption we have made about working in the Einstein's frame a few comments are in order. Let us recall that this frame is that one  at which the dilaton is coupled to the matter sector, in contrast with the Jordan frame where such a scalar field appears in the geometric sector, as a direct co-responsible for the gravity. From a mathematical point of view, the both frame are related merely by a conformal 
transformation. From the physical point of view, however, a longstanding debate has been made by many authors about the real physical meaning of these two versions of Brans-Dicke theory, raising the necessary question on which one is the physically relevant frame. Although the widespread usage of the Jordan frame in studying gravitational problems in different contexts \cite{faraoni} (and references therein), it is known that it faces drawbacks concerning stability \cite{faraoni1}, since within this frame the energy density of the dilaton is not bounded from below, what allows for an undesireable decaying toward lower and lower energy states, indefenitely. In \cite{faraoni}, the authors show that this feature implies a violation of the weak energy condition, with serious physical consequences that might jeopardize, for example, the propagation of gravitational waves. As they remark, this issue can be avoided when the Einstein frame is used instead of the Jordan one. These reasons support us in electing the Einstein frame as the framework where our analysis will evolve. Besides, we have additional two reasons for this choice: firstly, as in the Einstein frame the dilaton and graviton are indeed decoupled, it renders a more manageable system of differential equations. Secondly, the previous analysis using other topological defects to generate the bulk/brane structure were performed in this frame. Therefore, in order to be able to compare results and outputs, we shall present our analysis in the same context.

The energy-momentum tensor is defined in terms of the matter lagrangian as 
\begin{equation}
\label{emt}
T_{\mu\nu}=2\frac{\delta {\cal L}}{\delta g^{\mu\nu}}-g_{\mu\nu}{\cal L}. 
\end{equation}
For the global monopole the corresponding lagrangian density is
\begin{equation}
\label{GMlag}
{\cal L}=\frac{1}{2}\partial_{\mu}\phi^{a} \partial^{\mu}\phi^{a}-\frac{1}{4} \lambda (\phi^{a}\phi^{a}-\eta^2)^2\ ,
\end{equation}
which exhibits the symmetry breaking of the $SO(3)$ to $U(1)$ groups. In (\ref{GMlag}) the parameter $\lambda$ is a positive coupling constant and $\eta$ is the symmetry breaking scale. The $SO(3)$-symmetric Higgs field $\phi^a$ is an isotriplet of scalar fields which is assumed to have the form of the well-known hedgehog {\it Ansatz}, 
\begin{equation}
\label{campo}
\phi^{a}=\eta h(r) \frac{x^{a}}{r}\ .
\end{equation}
With the index $a=1,2,3$ and
\begin{equation}
\label{versor}
\hat{x}^{a}=\{\sin \theta \cos \phi, \sin \theta \sin \phi, \cos \theta \}.
\end{equation}

Using the expressions above we get the non-vanishing components of the energy-momentum tensor:
\begin{align}
\label{comp0}
&T_{0}^{0}=T_{z_1}^{z_1}=\;...\;=T_{z_p}^{z_p}=\frac{1}{2}\eta^2h'^{2}e^{-2\alpha}+e^{-2\beta}\eta^2 h^2+\frac{\lambda \eta^4}{4}(h^2-1)^2,\\
\label{comp1}
&T_{1}^{1}=-\frac{1}{2}\eta^2h'^{2}e^{-2\alpha}+e^{-2\beta}\eta^2 h^2+\frac{\lambda \eta^4}{4}(h^2-1)^2,\\
\label{comp2}
&T_{2}^{2}=T_{3}^{3}=\frac{1}{2}\eta^2h'^{2}e^{-2\alpha}+\frac{\lambda \eta^4}{4}(h^2-1)^2.
\end{align}
as well as its trace
\begin{equation}
T=\frac{(2+p)}{2}\eta^2h'^{2}e^{-2\alpha}+(2+p)e^{-2\beta}\eta^2 h^2+\frac{4+p}{4}\lambda \eta^4(h^2-1)^2.  
\end{equation}

The only non-trivial equation emerging from (\ref{consL}) is obtained for $\nu=1$:
\begin{eqnarray}
\label{consL1}
&&-\eta^2 h'h''e^{-2\alpha}-(\gamma'+2\beta'-\alpha'+p\sigma')\eta^2 h'^2 e^{-2\alpha}+2hh'\eta^2e^{-2\beta}+\lambda \eta^4 (h^2-1)hh'=\nonumber\\
&=&\tilde{\alpha}\left[\frac{(2+p)}{2}\eta^2 h'^2 e^{-2\alpha}+(2+p)e^{-2\beta}\eta^2h^2+\frac{(4+p)}{4}\lambda \eta^4 (h^2-1)^2\right]\varphi'. \label{eis}
\end{eqnarray}
In order to better manage the equation above we can assume we are performing our study far from the monopole's core, where the Higgs field behaves as $h\sim 1$. According to (\ref{consL1}) this condition is achieved if $e^{-2\beta} \rightarrow 0$ outside the monopole's core. Besides, the components of the energy-momentum (\ref{comp0})-(\ref{comp2}) outside the core assumes a much simpler form given by 
\begin{equation}
\label{emt1}
T_{\mu}^{\nu}\approx\textrm{diag}\left[e^{-2\beta}\eta^2\left(1,\;1,\;0,\;0,\;1,...,1\right)\right]\ .
\end{equation} The notation used here, including particularly the last result, shall be useful to the next section. When investigating the braneworld sum rules we shall make use of another notation.  

\subsection{Solving the field equations}

From (\ref{feq1}) and (\ref{feq2}) we obtain the following set of dynamical equations
\begin{align}
\label{eq1} 
&\left[\varphi''+(\gamma'-\alpha'+2\beta'+p\sigma')\varphi'\right]e^{-2\alpha}=4 \pi {\cal G} \tilde{\alpha} (2+p) e^{-2\beta}\eta^2,\\
\label{eq2}
&\gamma''+\gamma'(\gamma'-\alpha'+2\beta'+p\sigma')=0,\\
\label{eq3}
&\gamma''+2\beta''+p\sigma''+2\beta'^2-\alpha'(\gamma'+2\beta'+p\sigma')+\gamma'^2+p\sigma'^2=-2\varphi'^2,\\
\label{eq4}
&\left[\beta''+\beta'(2\beta'-\alpha'+\gamma'+p\sigma')\right]e^{-2(\alpha-\beta)}-1=-8 \pi {\cal G}\eta^2,\\  
\label{eq5}
&\sigma''+\sigma'(\gamma'-\alpha'+2\beta'+p\sigma')=0.
\end{align}

As a first attempt let us adopt a radial coordinate such that we have a type of higher dimensional version of the ``harmonic coordinate" introduced in \cite{kirill}, which brings an enormous simplification of gravitational problems involving scalar fields. In a scalar field model without a potential this choice is characterized by a radial coordinate $\tilde{u}$ satisfying $\Box \tilde{u}=0$ in which the condition $\gamma-\alpha+2\beta+p\sigma=0$ holds. Besides, we assume the condition $\gamma=\sigma$, necessary to construct the braneworld model we are interested in\footnote{This choice shall bring Lorentz invariance to the non compact dimensions performing (part of) the brane.}. Thus, we are left with the set of field equations below    
\begin{align}
\label{EQ1} 
&\varphi''=4 \pi {\cal G} \tilde{\alpha} (2+p) e^{2(\alpha-\beta)}\eta^2,\\
\label{EQ2}
&\gamma''=0,\\
\label{EQ3}
&2\beta''+2\beta'^2-\alpha'^2+(p+1)\gamma'^2=-2\varphi'^2,\\
\label{EQ4}
&\beta''e^{-2(\alpha-\beta)}-1=-8 \pi {\cal G}\eta^2.
\end{align}
The set of equations above can be analytically integrated, as we shall now see in detail. The integration of (\ref{EQ2}) is quite straightforward, providing the expression 
\begin{equation}
\label{gamma} 
\gamma(r)=\sigma(r)=ar+b,
\end{equation}
where $a$ and $b$ are integration constants. Next, let us examine the equations (\ref{EQ1}) and (\ref{EQ2}) which shall give us a very useful relation. Notice that the equation (\ref{EQ4}) may be rewritten as
\begin{equation}
\label{beta} 
\beta''(r)=(1-8 \pi {\cal G}\eta^2)e^{2(\alpha-\beta)}.
\end{equation}
This lead us to compare it with (\ref{EQ1}) and quickly set the interesting relation below
\begin{equation}
\label{phibeta} 
\varphi''(r)=\frac{4 \pi {\cal G} \tilde{\alpha} (2+p)\eta^2}{1-8\pi G \eta^2}\beta''(r). 
\end{equation}
By integrating twice this equation we are left with
\begin{equation}
\label{phibeta1} 
\varphi(r)=\frac{4 \pi {\cal G} \tilde{\alpha} (2+p)\eta^2}{1-8\pi G \eta^2}\beta(r)+B_0 r+\varphi_0. 
\end{equation}
For convenience, we can fix to zero the first integration constant appearing above. The second one, $\varphi_0$, shall be necessarily non-zero in order that the limit of the General Relativity (GR), namely $\varphi\rightarrow \textrm{const.}$ when $\tilde{\alpha}\rightarrow0$, may be properly recovered. Now let us define the constant factor multiplying $\beta(r)$ as
\begin{equation}
\label{constD} 
D\equiv \frac{4 \pi {\cal G} \tilde{\alpha} (2+p)\eta^2}{1-8\pi G \eta^2},
\end{equation}
so that (\ref{phibeta1}) will assume a simpler form 
\begin{equation}
\label{phibeta2} 
\varphi(r)=D\beta(r)+\varphi_0. 
\end{equation}
Now we are ready to address the remaining field equation. The relation above allows us to express (\ref{EQ3}) as a purely differential equation for $\beta(r)$. Using (\ref{gamma}) and (\ref{phibeta2}) in (\ref{EQ3}) and then collecting like terms we arrive at
\begin{equation}
\label{beta1} 
 2\beta''+2(D^2-1)\beta'^2-4a(p+1)\beta'-p(p+1)a^2=0.
\end{equation}
The next step, which shall make easier our job, is to adopt the change of variable $u\equiv ar+b$. So, by rewriting (\ref{beta1}) in terms of the new radial coordinate $u$, we get 
\begin{equation}
\label{beta1} 
 \ddot{\beta}(u)+(D^2-1)\dot{\beta}^2(u)-2(p+1)\dot{\beta}(u)-\frac{p(p+1)}{2}=0,
\end{equation}
where the dot denotes derivative with respect to $u$. A new change of variable can be performed, namely
\begin{equation}
\label{change}
\dot{\beta}\equiv f,
\end{equation}
which leads (\ref{beta1}) to the form
\begin{equation}
\label{beta2}
\dot{f}+(D^2-1)f^2-2(p+1)f-\tilde{A}=0,
\end{equation}
where $\tilde{A}\equiv \frac{p(p+1)}{2}$. It is easy to see that (\ref{beta2}) can be rewritten as
\begin{equation}
\label{beta3}
\dot{f}+(D^2-1)\left[f-\frac{(p+1)}{(D^2-1)}\right]^2-\frac{(p+1)^2}{(D^2-1)^2}-\tilde{A}=0.
\end{equation}
The equation above can be put in a nicer appearance through a further change of variable, in which we define
\begin{equation}
\label{change1}
F\equiv f-\frac{(p+1)}{(D^2-1)},
\end{equation}
and
\begin{equation}
\label{Btilde}
\tilde{B}\equiv \frac{(p+1)^2}{(D^2-1)^2}+\tilde{A}.
\end{equation}
Therefore, Eq. (\ref{beta3}) turns out to be
\begin{equation}
\label{beta4}
F'(u)+(D^2-1)F^2(u)-\tilde{B}=0.
\end{equation}
Rearranging this last expression we have
\begin{equation}
\label{beta5} 
\frac{dF}{\tilde{B}+(1-D^2)F^2}=du, 
\end{equation}
which is now ready to be integrated\footnote{Let us recall that: $\int \frac{dx}{ax^2+bx+c}=\frac{2}{\sqrt{4ac-b^2}}\tan^{-1}\left(\frac{2ax+b}{\sqrt{4ac-b^2}}\right)+c$.}. After some algebraic manipulations we find
\begin{equation}
\label{beta6} 
F=\frac{\tilde{B}(1-D^2)}{1-D^2}\tan \left[\sqrt{\tilde{B}(1-D^2)}u\right].
\end{equation}

Now we can move backwards along our multiple redefinition of variables. Using (\ref{change}) in (\ref{beta6}) we have
\begin{equation}
\label{beta7} 
f-\frac{(p+1)}{(D^2-1)}=\frac{\sqrt{\tilde{B}(1-D^2)}}{1-D^2}\tan \left[\sqrt{\tilde{B}(1-D^2)}u\right].
\end{equation}
Next, we use (\ref{change}) in (\ref{beta1}) which gives
\begin{equation}
\label{beta8} 
\dot{\beta}-\frac{(p+1)}{D^2-1}=\frac{\tilde{B}(1-D^2)}{1-D^2}\tan \left[\sqrt{\tilde{B}(1-D^2)}u\right].
\end{equation}
Integrating Eq. (\ref{beta8}) and resuming the usage of the original radial coordinate, $r$, we are left with the final form for $\beta(r)$:
\begin{equation}
\label{solBeta} 
\beta(r)=-\frac{1}{1-D^2}\ln \left\{\cos \left[\sqrt{\tilde{B}(1-D^2)}(ar+b)\right]\right\}-\frac{(p+1)}{(1-D^2)}(ar+b). 
\end{equation}

Let us recall that the metric function $\alpha(r)$ relates to $\gamma(r)$ and $\beta(r)$ by means of the coordinate condition we adopted, $\gamma-\alpha+2\beta+p\sigma=0$, which becomes 
\begin{equation}
\label{coord} 
(p+1)\gamma+2\beta=\alpha, 
\end{equation}
when we take into account the previously assumed condition $\gamma=\sigma$. 
Therefore, the exact solutions both for the dilaton and the metric functions, considering a generic number of extra dimensions $p$ read:
\begin{align}
\label{S1} 
&\varphi(r)=D\beta(r)+\varphi_0,\\
\label{S2}
&\gamma(r)=\sigma(r)= ar+b,\\
\label{S3}
&\beta(r)=-\frac{1}{1-D^2}\ln \left\{ \cos \left[\sqrt{\tilde{B}(1-D^2)}(ar+b)\right] \right\}-\frac{(p+1)(ar+b)}{(1-D^2)},\\
\label{S4}
&\alpha(r)=(p+1)(ar+b)-\frac{2}{1-D^2}\ln \left\{ \cos \left[\sqrt{\tilde{B}(1-D^2)}(ar+b)\right] \right\}-\frac{2(p+1)(ar+b)}{(1-D^2)}.
\end{align}
We remark that such an exact result is an entirely new in the literature.

Considering a realistic braneworld as modeling our universe we are led to fix $p=3$. Besides, as it is expected that the scale of symmetry breaking is quite below the Planck scale it is reasonable to assume ${\cal G}\eta^2\ll 1$, which makes $D^2\approx0$. Strictly speaking, there is no need for $\eta$ to respect the usual four dimensional grand unification scale, but we shall keep this order of magnitude to fix ideas. These further assumptions implies $\tilde{B}\approx 22$ and thus
\begin{align}
\label{S1} 
&\varphi(r)=D\beta(r)+\varphi_0,\\
\label{S2}
&\gamma(r)=\sigma(r)=ar+b,\\
\label{S3}
&\beta(r)=-\ln \left\{ \cos \left[\sqrt{22}(ar+b)\right] \right\}-4(ar+b),\\
\label{S4}
&\alpha(r)=-4(ar+b)-2\ln \left\{ \cos \left[\sqrt{22}(ar+b)\right] \right\}.
\end{align}

\subsection{Discussion}

We shall present a qualitative analysis/description of the obtained result in the light of the braneworld model it may engender. First we call attention to the fact that, provided well behaved quantum fluctuations, the size of the compact dimensions on the brane is fixed by the dynamical equations \cite{CE}. This size, however, needs to respect observational constraints \cite{expe}. There is a well known seesaw-like behavior more or less ubiquitous in higher dimensional braneworld models whose bulk/brane structure comes from a topological defect. This behavior is related to well defined extra dimensions on the brane and their impact on the hierarchy problem \cite{global,el} in the hybrid compactification scheme. Usually when the compact on-brane extra dimensions have a size compatible with experiments, the hierarchy problem is worsted and vice-versa. 

For the case at hands, it can be promptly verified, from Eqs. (\ref{S3}) and (\ref{S2}), that this seesaw behavior is reproduced here for $a>0$. That is, supposing the brane surface at say $r\rightarrow r^*$, then $\sigma(r\rightarrow r^*)$ and $\beta(r\rightarrow r^*)$ are such that the seesaw is present. Notice, however, that this case cannot fulfill the requirement stressed right after Eq. (\ref{consL1}), necessary to achieve $h\sim 1$ outside the monopole's core. Nevertheless the case $a<0$ ($b<0$) represents a interesting case, where it is possible to envisage a brane surface\footnote{Obviously, assuming that the brane surface is placed at a finite proper radial distance far away from the monopole's core.} respecting $|a|r^*<(\pi/2+|b|)/\sqrt{22}$, such that the size of the compact on brane dimensions may be keep small enough, along with a resulting warp factor which could serve, at least, to mitigate the hierarchy problem, thus preventing the model from a seesaw effect. Of course, the brane surface position must be fixed, for instance, by an internal solution and adequate junction conditions. In this case, however, at least part of the model inputs performed in Section II needs eventual reconsideration, resulting in a far from trivial problem. However, this is a different approach which is beyond the scope of this paper.        

The resulting line element for the solution found is clearly intricate and needs specific tools to make its physical content clear. In this regard we have computed the Kretschmann scalar for the spacetime given by Eqs. (\ref{S1}-\ref{S4}) and the result indicates a possibility of a spacetime completely free of singularities for suitable choice of the integration constants. By setting $b=0$ we shall have the following result:
\begin{eqnarray}
\label{Kret} 
K&=&4\cos^2(\sqrt{22}a r)\left[304\sqrt{22}a^4\cos^3(\sqrt{22}a r)\sin(\sqrt{22}a r)-1402 a^4\cos^4(\sqrt{22}a r)+12(198a^2+1)a^2\cos^2(\sqrt{22}ar)\right.\nonumber\\&+&16(1-22a^2)\sqrt{22}a^2\cos(\sqrt{22}a r)\sin(\sqrt{22}a r)+(1-22a^2) \left.\right]e^{16ar}.
\end{eqnarray}
By choosing a small enough and negative $a$, it is easy to see that exponencial factor present above shall dominate over the combination of trigonometric functions for large $r$, so that $K \rightarrow 0$ as $r\rightarrow \infty$, certifying the regularity of this spacetime, as illustrated in the plot below:

\begin{figure}[h]
\begin{center}
{\includegraphics[width=8.5cm,height=7.0 cm]{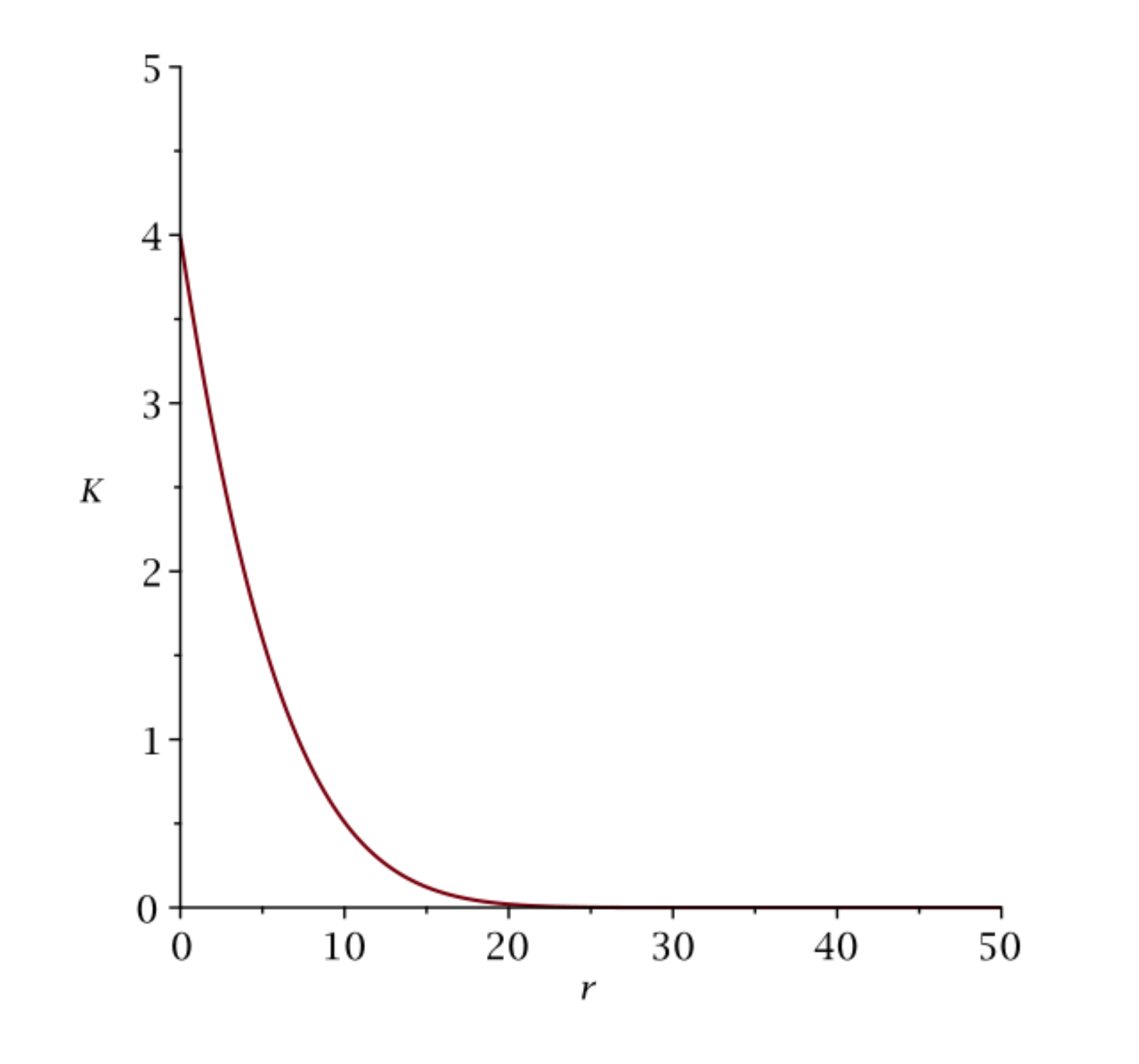}}
\end{center}
{\caption{Behavior of the Kretschmann scalar along the radial distance when $a=-10^{-3}$. It shows clearly the absence of singularity in the spacetime we are considering when a suitable choice of the integration constants are made.}}
\label{f1}
\end{figure} 

This is a relevant output, since singularities appearing in gravitational solutions coming from global defects occurs more often than never. Nevertheless, although the regularity of the obtained spacetime is an important advantage of the present model, the functional form of the dilatonic field constitutes an inconvenient issue to the scenario under study. In Einstein frame the dilatonic field couples to the ordinary matter, so that we hope it has a well-behaved profile, even assymptotically. However, it may be readily verified that the dilatonic field is not everywhere well defined in its domain, due to the presence of a logarithm divergence of the scalar field as $r$ approaches the infinity. Even though this behavior be not desirable, we reinforce the role of the Brans-Dicke field making evident the need of a more involved setup. For example, a possible way out would be considering a multiple brane scenario along with a nontrivial radial extra dimension topology in order to bypass this problem. The branes (or some of the branes) could, as usual, be placed at singular points of orbifolds, cutting off problematic radial regions. We shall leave this analysis for the next section, not by studying an specific example (whose complications may be easily underestimated), but otherwise by means of the so-called braneworld sum rules. This formalism correspond to a set of consistency rules which comprises more general braneworld scenarios, like the multiple branes one mentioned above.   

\section{Sum rules analysis}

The problem approached in this paper pointed, in the previous section, to the necessity of a more involved configuration, in order to avoid completely the ill defined regions to the dilatonic field along the radial extra dimension. However, a direct approach, putting two (or several) branes in the spacetime within a radial nontrivial topology, aiming to solve the field equations, seems to be out of order, given the complexity of this endeavor. We can, on the other hand, investigate consistency conditions to be respected by such a braneworld model, possessing at least two branes and orbifold-like radial extra dimension.    

The braneworld sum rules where first proposed in Ref. \cite{KLG} applied to the five-dimensional paradigmatic braneworld \cite{RSI} and short after generalized to a plethora of possible scenarios \cite{LLL}. The basic idea is to make use of the compact internal space (the radial extra dimension in our case) along with an ingenious manipulation of the bulk/brane geometric quantities, in order to extract physical information about a given model without properly solving the field equations. The braneworld sum rules provide a general formalism to extract precise physical information about the (braneworld) system even in the case of intractable gravitational equations performing, in this context, a precise tool investigating the physical outputs of the braneworld model in the absence of analytical solution. The sum rules method was previously applied in the context of Brans-Dicke gravity in Ref. \cite{13}, where it was shown that the dilatonic gravity allows for smooth branes in five dimensions and do not needs negative brane tensions as well. All the results obtained so far, in the scalar-tensor formalism, formulate the sum rules in the Jordan's frame. Our model, however, was built in the Einstein's frame and since the very existence of the branes lead to a conformal invariance breaking, there is no general argument pointing to the possibility of using here the results obtained in the Jordan's frame {\it a priori}. Therefore we shall recompute the braneworld sum rules in the Einstein's frame.  

Since here we shall deal with bulk, brane and internal space in a systematic manner, the notation used in this section must differ from the used so far. Therefore, we shall fix the notation as follows: before to particularize the formulation to our case, we denote by $\mathcal{D}=p+4$ the number of bulk dimensions, by $\tilde{p}+1$ the number brane dimensions and by $\mathcal{D}-\tilde{p}-1$ the number of dimensions of the internal space. The indices for the whole bulk shall be denoted by capital Latin letters, while for the brane dimensions we reserve Greek letters. Latin small letters stands for the internal space dimensions. Besides, while the bulk scalar of curvature is denoted by $R$, $\bar{\bar{R}}$ denotes the brane scalar of curvature, while $\tilde{\tilde{R}}$ stands for its internal space counterpart. From this general perspective the line element may be denoted by 
\begin{equation}
ds^2=G_{AB}dx^A dx^B=W^2(r)g_{\alpha\beta}dx^\alpha dx^\beta+g_{ab}dr^a dr^b. \label{AS}
\end{equation} Regarding the notation here used, notice that the brane coordinates may also encompass compact extra dimensions (this is indeed our case of interest) while the transverse dimensions to the brane are reserved to the internal space.    

A relatively simple manipulation of geometric quantities (see Ref. \cite{LLL} for details) allows one to relate the scalar of curvatures and partial traces (see below) with a total derivative in the internal space as 
\begin{eqnarray}
\partial\cdot(W^\chi(r) \partial W(r))=\frac{W^{\chi+1}}{\tilde{p}(\tilde{p}+1)}\Big[\chi(\bar{\bar{R}}W^{-2}(r)-R^{\mu}_{\mu})+(\tilde{p}-\chi)(\tilde{\tilde{R}}-R^m_m)\Big],\label{guas}
\end{eqnarray} where $\chi$ is a free parameter and the partial traces are defined as $R^{\mu}_{\mu}=W^{-2}(r)g^{\mu\nu}R_{\mu\nu}$ and $R^m_m=g^{mn}R_{mn}$, such as $R=R^{\mu}_\mu+R^m_m$. When the specific model/gravitational theory is implemented via partial traces and an integral over the internal compact space is performed, a one-parameter ($\chi$) family of consistency conditions arise. These conditions are the sum rules.    

In the new notation Eq. (\ref{feq2}) is given by 
\begin{equation}
\label{Nfeq1}
R_{N}^{M}=2 \partial^{M}\varphi \partial_{N}\varphi + 8\pi {\cal G}\left(T_{N}^{M}-\frac{\delta_{N}^{M}T}{p+2}\right),
\end{equation} from which the partial traces can be read. Taking into account that the Brans-Dicke field has no dynamics on the brane and recalling that $g^{\mu\nu}g_{\mu\nu}=\tilde{p}+1$ we have 
\begin{equation}
R^{\mu}_\mu=8\pi {\cal G}T^{\mu}_\mu-\frac{8\pi {\cal G}(\tilde{p}+1)}{p+2}T, \label{ol}
\end{equation} 
\begin{equation}
R^m _m=2g^{mn}\partial_m\varphi\partial_n\varphi+8\pi {\cal G}\left(T^m_m-\frac{\mathcal{D}-\tilde{p}-1}{p+2}T\right), \label{la}
\end{equation} with $T^\mu_\mu$ ($T^m_m$) defined in a manner akin to $R^{\mu}_\mu$ ($R^m_ m$). Inserting Eqs. (\ref{ol}) and (\ref{la}) into Eq. (\ref{guas}) we have 
\begin{eqnarray}
\partial\cdot(W^\chi(r) \partial W(r))&=&\left.\frac{W^{\chi+1}}{\tilde{p}(\tilde{p}+1)}\Bigg[\chi\Bigg(\bar{\bar{R}}W^{-2}-8\pi {\cal G}T^\mu_\mu+\frac{8\pi {\cal G}(\tilde{p}+1)}{p+2}T\Bigg) \right.\nonumber\\&+&\left. (\tilde{p}-\chi) \Bigg(\tilde{\tilde{R}}-2g^{mn}\partial_m\varphi\partial_n\varphi-8\pi {\cal G}\left(T^m_m-\frac{\mathcal{D}-\tilde{p}-1}{p+2}T\right)\Bigg)\Bigg].\right. \label{fim1}
\end{eqnarray} Working out the right hand side of Eq. (\ref{fim1}), by taking into account that $T=T^{\mu}_{\mu}+T^m_m$, we arrive at 
\begin{eqnarray}
\partial\cdot(W^\chi(r) \partial W(r))&=&\left.\frac{W^{\chi+1}}{\tilde{p}(\tilde{p}+1)}\Bigg\{\chi\bar{\bar{R}}W^{-2}+8\pi {\cal G}T^\mu_\mu\Bigg[-\chi+\frac{(\tilde{p}+1)\chi}{p+2}+\frac{(\tilde{p}-\chi)(\mathcal{D}-\tilde{p}-1)}{p+2}\Bigg]+(\tilde{p}-\chi)\tilde{\tilde{R}}\right.\nonumber\\&-&\left. 2(\tilde{p}-\chi)g^{mn}\partial_m\varphi\partial_n\varphi+8\pi {\cal G}T^m_m \Bigg[ \frac{\chi(\tilde{p}+1)}{p+2}+(\tilde{p}-\chi)\Bigg(-1+\frac{\mathcal{D}-\tilde{p}-1}{p+2}\Bigg)\Bigg]\Bigg\}.\right.
\end{eqnarray}  

Now we are in position to apply the formalism to a braneworld scenario motivated by the investigations performed in the previous section. We shall first particularize the dimensions and in a second moment entering the global monopole stress tensor. The case investigated in the previous section pointed to a bulk with $\mathcal{D}=7$ (and hence $p=3$) and $\tilde{p}=5$. Therefore the brane is six-dimensional and the internal space $(\mathcal{D}-\tilde{p}-1)$ has 1 dimension. Notice that for the codimension one case $\tilde{\tilde{R}}=0$. Therefore, denoting $\partial_r \varphi=\varphi'$, we arrive at 
\begin{eqnarray}
\partial\cdot(W^\chi(r) \partial W(r))=\frac{W^{\chi+1}}{30}\Bigg\{\chi\bar{\bar{R}}W^{-2}+2(5-\chi)g^{rr}(r)(\varphi')^2+8\pi {\cal G}\Big[T^{\mu}_\mu+2(\chi-2)T^m_m\Big]\Bigg\}.\label{ns}
\end{eqnarray} 

The stress tensor partial traces occurring in Eq. (\ref{ns}) are given by    
\begin{eqnarray}
T^{\mu}_\mu=-6\sum_i T_5^{(i)}\delta(r-r_i)+W^{-2}g^{\mu\nu}\tau_{\mu\nu}, \nonumber\\
T^m_m=g^{mn}\tau_{mn}, \label{nss}
\end{eqnarray} where $T_5^{(i)}$ stands for the $i^{th}-$brane tension. Notice that these partial traces comes from a generic bulk stress tensor, without cosmological constant, allowing for several branes placed at $r_1, r_2, \cdots$. The monopole stress tensor, denoted here by $\tau_{MN}$, is the usual one
\begin{equation}
\tau_{MN}=\partial_M\phi^a\partial_N\phi^a-\frac{1}{2}G_{MN}G^{KL}\partial_K\phi^a\partial_L\phi^a+G_{MN}V(\phi^a). \label{o} 
\end{equation} Hence the partial traces read
\begin{eqnarray}
\tau^\mu_\mu=W^{-2}g^{\mu\nu}\tau_{\mu\nu}=-2W^{-2}(r)g^{\mu\nu}\partial_{\mu}\phi^a\partial_{\nu}\phi^a-3g^{rr}(r)(\phi'^a)^2+6V(\phi^a),\nonumber\\
\tau^m_m=T^m_m=-\frac{1}{2}W^{-2}(r)g^{\mu\nu}\partial_{\mu}\phi^a\partial_{\nu}\phi^a+\frac{1}{2}g^{rr}(\phi'^a)^2+V(\phi^a).\label{g}, 
\end{eqnarray} where the first two terms of the partial trace survives due to the monopole hedgehog configuration. Inserting back Eqs. (\ref{g}) into (\ref{nss}) and finally into (\ref{ns}) we have 
\begin{eqnarray}
\partial\cdot(W^\chi(r) \partial W(r))&=&\left.\frac{W^{\chi+1}}{30}\Bigg\{\chi\bar{\bar{R}}W^{-2}+2(\chi-5)g^{rr}(r)\Big[4\pi {\cal G}(\phi'^a)^2-(\varphi')^2\Big]\right.\nonumber\\&+&\left. 16\pi {\cal G}(\chi+1)V(\phi)-8\pi{\cal G}\chi W^{-2}(r)\partial^\mu\phi^a \partial_\mu \phi^a-48\pi {\cal G}\sum_iT_5^{(i)}\delta(r-r_i) \Bigg\}.\right. \label{qua}
\end{eqnarray} 

There are some important remarks in order to carry the calculations a little further. To achieve a more concrete braneworld scenario we can think of two branes placed at the singular points of a $\mathbb{Z}_2$ orbifold. Within this case we certainly may avoid the radial problematic regions present in the last section. Moreover, this context is the simplest realization of a compact internal space for which $\oint \partial\cdot(W^\chi(r) \partial W(r))=0$, leading to the one parameter family of consistency conditions. Hence, integrating over the internal space we have 
\begin{eqnarray}
	\oint W^{\chi+1}\Bigg\{\chi \bar{\bar{R}}W^{-2}+2(5-\chi)g^{rr}(r)\Big[4\pi {\cal G}(\phi'^a)^2-(\varphi')^2\Big]&+&\left. 16\pi {\cal G}(\chi+1)V(\phi^a)-8\pi{\cal G}\chi W^{-2}(r)\partial^\mu\phi^a \partial_\mu \phi^a\Bigg\}\right.\nonumber\\&=&\left.48\pi {\cal G}\Big(T_5^{(1)}W^{\chi+1}(r_1)+T_5^{(2)}W^{\chi+1}(r_2)\Big).\right.\label{fd}
\end{eqnarray} Eq. (\ref{fd}) may be recast in a slightly more informative fashion by means of a simplification. In trying to reproduce our Universe in large scales it is pretty safe to set $\bar{\bar{R}}\sim 0$ and, therefore, Eq. (\ref{fd}) reads \begin{eqnarray}
\oint  W^{\chi+1}\Bigg[\frac{(5-\chi)g^{rr}(r)}{4\pi {\cal G}}\Big[4\pi {\cal G}(\phi'^a)^2-(\varphi')^2\Big]&-&\left.\chi W^{-2}(r)\partial^\mu \phi^a \partial_\mu \phi^a+2(\chi+1)V(\phi^a)\Bigg]\right.\nonumber\\&=&\left. 6\Bigg\{T_5^{(1)}W^{\chi+1}(r_1)+T_5^{(2)}W^{\chi+1}(r_2)\Bigg\}.\right.\label{ff}
\end{eqnarray} Now the choice $\chi=-1$ is particularly illustrative since it eliminates most warp factor contributions and the contribution coming from the potential as well. Within this choice it is fairly simple to see that the monopole Higgs field dynamics may counterbalance the Brans-Dicke sign contribution in a way that no negative brane tension is needed. This aspect is in fact relevant, since in codimension one braneworld models, the brane tension is directly proportional to the effective four-dimensional Newtonian constant \cite{SSM}. 

\section{Conclusion}

In this paper we study an alternative braneworld model in a twofold approach. Firstly, we construct a framework where a global monopole is used to generate the bulk-brane setup, within a Brans-Dicke gravity, the simplest gravitational theory encompassing the dilatonic field. For such a system we derive and solve exactly the dynamical equations, obtaining a model in which a seven dimensional bulk accommodates a brane endowed with four non-compact dimensions and two (on-brane) compactified ones, obeying the hybrid compactification scheme. In this scope, the radial coordinate plays the role of the transverse dimension. Next it is analyzed the viability of the obtained braneworld scenario and how useful could it be in addressing the so-called hierarchy problem. As we have shown, this aim was successfully achieved in our model. Furthermore, we discussed at which extent the solution can be used in order to avoid the seesaw-like effect, a typical behavior characterizing braneworld models based on topological defects. We found that for a proper choice of the integration constants present in the solution the seesaw issue may be avoided, providing an auspicious hybrid compactification scenario. However, while the resulting spacetime is promising and everywhere regular (as indicated by the profile of the Kretschmann scalar), the Brans-Dicke scalar field evinces an asymptotic undesired behavior as the radial (transverse) dimension approaches the infinite. 

Motivated by this problem concerning the dilatonic field behavior along the radial direction we performed an analysis of possible model extensions in light of the sum rules formalism. This approach reveals us that, despite the additional terms coming from the Brans-Dicke field, it is possible to appreciate a more involved model containing, for instance, two positive tension branes placed at the singular points of a radial orbifold.

We finalize by stressing that the full model, whose brane surface position is stabilized by the solution inside the monopole’s core along with appropriate junction conditions at the brane surface (or eventually by another stabilization mechanism), is currently under investigation, but the usual approach has been shown quite non-trivial so far. This possibility may be explored in a future opportunity.

\begin{acknowledgments}
JMHS thanks to CNPq (Grant no. 303561/2018-1) for partial financial support.
\end{acknowledgments}

\end{document}